\documentclass[pre,superscriptaddress,twocolumn,a4paper,showpacs]{revtex4-1}
\usepackage{amssymb}
\usepackage{amsmath}
\usepackage{wasysym}
\usepackage{latexsym}
\usepackage{graphicx}
\usepackage{epsfig}
\usepackage{subfigure}
\usepackage{float}
\begin{document}

\title{Agent-based models of collective intelligence }

\author{Sandro M. Reia}
\affiliation{Instituto de F\'{\i}sica de S\~ao Carlos,
  Universidade de S\~ao Paulo,
  Caixa Postal 369, 13560-970 S\~ao Carlos, S\~ao Paulo, Brazil}

\author{Andr\'e C.  Amado}
\affiliation{Departamento de F\'{\i}sica, Universidade Federal de Pernambuco,  50670-901, Recife, PE, Brazil}

\author{Jos\'e F.  Fontanari}
\affiliation{Instituto de F\'{\i}sica de S\~ao Carlos,
  Universidade de S\~ao Paulo,
  Caixa Postal 369, 13560-970 S\~ao Carlos, S\~ao Paulo, Brazil}

\begin{abstract}
Collective or group intelligence is manifested in the fact that a team of cooperating agents can solve problems more efficiently than when those agents work in isolation. Although cooperation is, in general,  a successful problem solving strategy, it is not clear whether it merely  speeds up  the time to find the solution,  or whether it alters qualitatively  the statistical signature of the search for the solution.  Here we  review and offer insights on two agent-based models of distributed cooperative problem-solving systems, whose task is  to solve a cryptarithmetic puzzle. The first model is 
the imitative learning search  in which the agents exchange information on the quality of their partial solutions to the puzzle and  imitate the most successful agent in the group.  This scenario predicts a very poor performance in the case   imitation is too frequent or  the group is too large, a phenomenon  akin to  Groupthink of social psychology.
The second model is   the blackboard organization in which agents read and post hints on a public blackboard. This brainstorming scenario performs the best when there is a stringent  limit to the amount of information that is exhibited on the board. Both cooperative scenarios produce a substantial speed up of the time to solve the puzzle as compared with the situation where the agents work in isolation. The statistical signature of the search, however,  is the same as  that of  the independent search. 
\end{abstract}

\maketitle

\section{Introduction}\label{sec:intro}

The development of  software packages to tackle almost any technical problem we can think of, as well as of  friendly interfaces between people and computers (see, e.g., \cite{Fallani_18} in this issue), has contributed  to the tremendous increase of the productivity of today's  society,  thus ratifying the 1960s vision of the computer pioneer Doug Engelbart  that the human intellect could be augmented with the aid of computers \cite{Engelbart_62}. In fact, if  human intelligence is gauged by the capacity and speed to solve problems, which seems a sensible perspective for the scientific and technological  milieus,  then
it is difficult to disagree with Engelbart's rationale.

However, since the general intelligence of a person -- the  $g$ factor --  has strong genetic  (about 50\%) and developmental (the remaining  50\%) components  there is little  adults  can do to increase their intelligence (see \cite{Deary_01}  for a lucid discussion of these
controversial issues), thus limiting the growth of the most important element of the human-computer partnership. Although the natural and widely employed way to circumvent this limitation is to consider group work, only recently  psychologists have put forward evidence supporting a general collective intelligence factor -- the so-called $c$ factor -- that explains the  group  performance on a variety of  tasks \cite{Woolley_10}. Surprisingly, the $c$ factor  does not seem to  be strongly correlated with the average or maximum individual intelligence of group members. It is correlated instead with the average social sensitivity of the group members \cite{Woolley_10}. Therefore, we could, in principle, augment the group intelligence by properly selecting the group composition and internal organization.  

Actually, the study of the influence  of the organization and, in particular, of  intra-group communication patterns (i.e., who can communicate  with  whom) on  the problem-solving performance  of groups dates back at least  to the 1950s \cite{Bavelas_50,Leavitt_51,Guetzkow_55} (see \cite{Mason_12, Sebastian_17,Reia_17} for more recent contributions). Understanding this influence is, of course,  of immense value because
problem solving  (e.g., drug design, traffic engineering, software development) by task forces  represents a substantial portion  of the economy  of developed countries nowadays \cite{Page_07}.  

As argued above, we see organizational design as a by-product of the  study  of collective intelligence or, more specifically, of  distributed cooperative problem-solving systems. The key feature of these systems is that their members exchange information about their progress towards the completion of a goal \cite{Huberman_90,Clearwater_91}.
There are many common-sense assumptions in this field, e.g., that a group of cooperating individuals is more efficient than those same individuals working in isolation or that diversity  is always beneficial to the group performance, that were not fully scrutinized through  a  powerful (in the explanatory  sense)  analytical tool of physics, namely,  the  mathematical and computational modeling of complex phenomena using
 minimal models. A minimal  model  should exhibit a good balance between simplicity and realism and  should be successful in reducing  a complex collective phenomenon to its functional essence (see  \cite{Perlovsky_06,Felix_18} for use of this approach to elucidate the physics of mind).

In fact, despite the extensive  use of optimization heuristics inspired on cooperative systems, such as the particle swarm optimization algorithm \cite{Bonabeau_99} and the adaptive culture heuristic \cite{Kennedy_98,Fontanari_10}, to search for  optimal or near optimal solutions of combinatorial problems, we know little about the factors that make cooperation effective, as well as about the universal character (if any) of the quantitative improvements that results from it \cite{Clearwater_91}.  This is so  because  those heuristics and the problems they are set to solve are too complex to yield to a first-principle analysis. Here we review recent attempts to study distributed cooperative problem-solving systems using minimal models  following the research strategy set forth  by  Huberman in the late 1980s \cite{Huberman_90,Clearwater_91,Huberman_87}.
We consider two scenarios of distributed cooperative systems where  the goal of the agents is to solve a particular cryptarithmetic puzzle [see eq.\  \ref{DGR})], which is an illustrative example of the  class of constraint satisfaction problems that bases most studies on problem solving \cite{Newell_72}. 

In the first scenario we  endow the agents with the capacity to evaluate the goodness of the partial solutions of their peers and imitate the more successful agent in the group.  In the context of collective intelligence or global brains,  imitative learning is  probably the  most important  factor  as  neatly expresses  this quote by Bloom  ``Imitative learning acts like a synapse, allowing information to leap the gap from one creature to another'' \cite{Bloom_01}.  Since imitation  is central to the remarkable success of our species \cite{Blackmore_00, Rendell_10}
(see \cite{Kivinen_18} for a discussion of this issue),  we expect that this scenario  may be of relevance to the organization of real-world task groups  \cite{Lazer_07}. In fact, we found that if the agents are too propense to imitate their more successful peers or if the group is too large then the  group performance is catastrophic when compared with the baseline situation where  the
agents work independently of each other \cite{Fontanari_14, Fontanari_15}. This is similar to the classic Groupthink phenomenon of social psychology that occurs when everyone in a group starts thinking alike \cite{Janis_82}.  Avoiding this sort of  harmful effect is the task of organizational designers and we have  verified that two  rather natural interventions are partially effective to circumvent Groupthink, namely, 
decreasing the connectivity of the agents so as to delay the propagation of misleading information through the system \cite{Rodrigues_16}, and
allowing  diversity in the  agents'   propensities to imitate their peers \cite{Fontanari_16}. However, these interventions have an unwelcome side effect: the degradation of the optimal  performance of the group, which is achieved in the case of  fully connected homogeneous agents. 

The second scenario of distributed cooperative problem-solving systems  that we consider here is the blackboard organization that was introduced in the Artificial Intelligence  domain in the 1980s and is now part of the AI problem-solving toolkit \cite{Corkill_91}. In this organization, the agents read and write hints to a central blackboard that can be accessed by all members of the group. Hence the blackboard scenario  describes the  common  view  of task-forces  as  teams
of specialists exchanging ideas on possible approaches to solve a problem and displaying the promising suggestions in a public  blackboard. In this scenario, there is no need to assume that  the agents are capable of quantifying the goodness of their partial solutions to the cryptarithmetic puzzle as in the imitative learning  scenario (see, however, \cite{Fontanari_18}). To our knowledge, this was the first distributed cooperative problem-solving system studied using an agent-based minimal model  \cite{Clearwater_91}.  Contrary to the claims of that original study, however,  we found that the search using the blackboard organization  exhibits the same statistical  signature of the independent search \cite{Fontanari_18}.  Most unexpectedly,  we found that limiting the amount of information displayed on the board can markedly boost the performance of the group. This is an  original result that we offer in this review paper, which illustrates well the power of minimal models to reveal relevant hidden features of complex systems.

The rest of this short review paper is organized as follows. In Section \ref{sec:CP} we present  the particular cryptarithmetic problem that we use throughout the paper,  define  the costs associated to the digit-to-letter assignments that are necessary to implement  the imitative learning search strategy, and introduce the definition of hint  that is necessary for  implementing the blackboard organization.  In  Section \ref{sec:CC} we present our measure of the group performance, which is proportional to the time for an  agent in the group to find the solution and hence to halt the search. By dividing this time  by the size of the state space of the problem and multiplying it by the number of agents in the group we obtain the computational cost of the search, which is our performance measure.    In Section \ref{sec:imit}  we describe the imitative learning search strategy and discuss its performance on the cryptarithmetic puzzle for the simplest case where all agents interact with each other, i.e.,  the pattern of  communication  is a fully connected network. 
In Section \ref{sec:black}  we present the minimal model for  the blackboard organization  and show that the probability distribution of the computational costs is an exponential distribution as in the case of the independent search. The mean computational cost of the search for the standard blackboard organization, however, is about ten times lower than for the independent search. Also in that section, we introduce the limited space blackboard  scenario where the agents have to compete for space in the board. In this case, the computational cost can be reduced by another factor of ten using a judicious choice of the blackboard size. 
 Finally, Section \ref{sec:conc} is reserved to our concluding remarks and to advance some avenues for future research.

\section{The cryptarithmetic problem}\label{sec:CP}

Cryptarithmetic problems such as 
\begin{equation}\label{DGR}
DONALD + GERALD = ROBERT 
\end{equation}
are constraint satisfaction problems in which the task is to find unique digit-to-letter assignments  so that the integer numbers represented by the words add up correctly \cite{Averbach_80}. In the cryptarithmetic problem  (\ref{DGR}), there are $10!$ different digit-to-letter assignments, of which only one is the solution to the problem, namely, $A=4$, $B=3$, $D=5$, $E=9$, $G=1$, $L=8$, $N=6$, $O=2$, $R=7$, $T=0$ so that 
$DONALD = 526485$, $GERALD = 197485$ and $ROBERT = 723970$.
 In this paper we will focus only on the cryptarithmetic problem  (\ref{DGR})  because its state space is the largest possible  for this type of puzzle (a cryptarithmetic puzzle has at most $10$ different letters) and because it facilitates the  replication of our findings. Use of randomly generated cryptarithmetic puzzles as well as distinct optimization problems, such as finding the global maximum of NK-fitness landscapes \cite{Kauffman_87}, has yielded the same (qualitative) results  \cite{Fontanari_14, Fontanari_15}.
 
The imitative learning search strategy  that will be  discussed in Sect. \ref{sec:imit}  requires that we assign a cost to each digit-to-letter assignment, which  is viewed as a measure of the goodness of the answer represented by that assignment. A natural choice for the cost function is \cite{Abbasian_09}
\begin{equation}\label{cost}
c  = \left | ROBERT - \left ( DONALD + GERALD \right ) \right |.
\end{equation}
For example, the  digit-to-letter assignment $A=0$, $B=2$, $D=9$, $E=4$, $G=8$, $L=1$, $N=7$, $O=6$, $R=3$, $T=5$ yields $ROBERT = 362435$, $DONALD= 967019$ and $GERALD = 843019$ and the cost assigned to it  is $c = 1447603$. We should note that the cost value  (\ref{cost}) applies to all  digit-to-letter assignments  except those for which $R=0$, $D=0$ and $G=0$, which are 
 invalid assignments since
they violate the rule of the cryptarithmetic puzzles that an integer number should not have the digit $0$ at its leftmost position. Hence for
those assignments we fix an arbitrary large cost value, namely, $c = 10^8$, so that now they become valid assignments but   have the
highest  cost among all assignments. If the cost of a digit-to-letter assignment  is $c=0$ then it is the  solution to the cryptarithmetic problem.

Although we could easily think up clever alternatives to the  cost function   (\ref{cost}), we recall that our aim is not to design efficient algorithms to solve  cryptarithmetic problems  but to explore  cooperative strategies that improve the efficiency of group work \cite{Huberman_90}. In that sense, the chosen problem (\ref{DGR}) is  quite challenging in that it offers many  misleading clues -- local minima of the cost (\ref{cost}) and  wrong hints  -- which may  lure the search away from the solution.  However, as already mentioned, we stress that the main advantage of considering a specific problem is the easy to replicate  and verify our claims.

At this stage it is convenient to describe the minimal or elementary move  in the state space composed of the $10!$ possible digit-to-letter assignments. 
Starting from a particular digit-to-letter assignment, say, $A=0$, $B=2$, $D=9$, $E=4$, $G=8$, $L=1$, $N=7$, $O=6$, $R=3$, $T=5$ we choose two different letters  at random and  interchange the digits assigned to them. For example, if we pick letters $D$ and $T$  then the assignment  that results from the application of the elementary move is $A=0$, $B=2$, $D=5$, $E=4$, $G=8$, $L=1$, $N=7$, $O=6$, $R=3$, $T=9$. 
Any  two valid digit-to-letter assignments that are connected by the elementary move are said to be neighbor assignments.
Hence each digit-to-letter  assignment in the state space of problem (\ref{DGR}) has exactly 45 neighbors. Clearly, the repeated application of our elementary move allows us to explore  the entire state space of the  cryptarithmetic problem.
  We can check all assignments   and their neighbors to find the number of minima, i.e., those assignments that have a cost (strictly) lower than the cost of their neighbors.  We find that the cryptarithmetic problem (\ref{DGR}) has 102  minima in total: a single global minimum and 101 local minima. We recall, however,  that 
the existence and characteristics of the  local minima are strongly dependent  on the choices of the cost function and of the elementary move in the state space. 

An important feature of cryptarithmetic puzzles, which makes them  a testbed for cooperative strategies, is the existence of hints that may  hint on the suitability of a particular 
 digit-to-letter assignment. A hint is a  set of letters  in a same column  that add up correctly modulo $10$. For example, considering the third column (from left to right) of the problem  (\ref{DGR}) we have   $B  =  \epsilon +  N +R  $ where $\epsilon = 0,1$ and the sum 
is done modulo $10$. The case  $\epsilon =1$ accounts for the possibility that an $1$  is carried from the sum of the letters in the fourth column.  Of course, for the rightmost column  ($D+D=T$) the only possibility is   $\epsilon = 0$. For this column there are 9 different hints: ($D=1$, $T=2$),
($D=2$, $T=4$),  ($D=3$, $T=6$), etc. Each one of the columns $\epsilon + L + L =R$, $\epsilon + A + A=E$  and $\epsilon + O+E=O$ has 18 different hints (9 for $\epsilon = 0$ and 9 for $\epsilon = 1$), whereas columns $\epsilon + N + R = B$ and $\epsilon + D + G = R$ have 144  different hints (72 for $\epsilon = 0$ and 72 for $\epsilon = 1$) each. Hence there are a total of $351$ distinct hints but only $6$ of them yield the  solution of the puzzle  (\ref{DGR}). We note that we could further reduce  this number by eliminating hints in the leftmost column   such that 
$ \epsilon +  D + G > 9$  with the sum now done modulo 1. However, we choose not to implement this rule since what matters is that the total
number of hints is much smaller than the size of the state space and that the six correct hints are contemplated in our definition of hint.
In the blackboard organization, the communication between the agents is achieved by   posting and reading  hints in a public blackboard \cite{Clearwater_91}  and so there is no need to introduce  a cost for each digit-to-letter assignment (see
\cite{Fontanari_18} for a scheme where the  hints are displayed  together with the costs of the agents that posted them).

\section{The computational cost}\label{sec:CC}

The efficiency of the search for the solution of  the cryptarithmetic puzzle is measured by the computational cost that is defined as follows. Let us consider a  group  composed of $M$  agents so that  each agent is represented by a digit-to-letter assignment. The agents
explore the state space following a  search strategy that specifies the rules for updating their digit-to-letter assignments.  Each time a randomly chosen agent updates its digit-to-letter assignment we increment the time $t$ by the quantity $\Delta t = 1/M$, so  that during the increment from $t$ to $t+1$ exactly  $M$, not necessarily distinct, agents are updated.  The search ends when one of the agents finds the solution to the puzzle and we denote by $t^*$ the time when this happens. Since we expect that $t^*$ will increase with the size of the state space  ($10!$ for our cryptarithmetic puzzle) and that, at least for the independent search, it will decrease with the reciprocal of the number of agents, we  define the computational cost $C$ of the search as
\begin{equation}\label{CC}
C =  M t^*/10! 
\end{equation}
so that $C$ is on the order of 1 for the independent search, regardless of the group size.
 Next we  study the statistical properties of the  computational cost for two cooperative problem-solving scenarios, namely, the imitative learning search and  the blackboard organization.

\section{The imitative learning search}\label{sec:imit}

This search strategy is based on the presumption that the cost (\ref{cost}) offers a clue on the goodness of the digit-to-letter assignment so that
it may be advantageous to copy or imitate agents whose assignments have low cost. More pointedly, 
at time $t$, a randomly chosen   agent -- the target agent -- can choose between two actions. The first action,
which happens with probability $1-p$,  is the elementary or minimal move in the space space described in Section \ref{sec:CP}.
The second action, which happens with probability $p$, is the  imitation  of the model agent, which is the agent with the lowest cost digit-to-letter assignment in the group at  time $t$. To illustrate the copying process let us assume for the sake of concreteness that the target agent has the 
assignment   $A=0$, $B=2$, $D=9$, $E=4$, $G=8$, $L=1$, $N=7$, $O=6$, $R=3$, $T=5$ whose cost is $c = 1447603$ and the model agent has the assignment  $A=5$, $B=3$, $D=9$, $E=4$, $G=8$, $L=1$, $N=6$, $O=2$, $R=7$, $T=0$ whose cost is $c = 1050568$. In the copying process the target agent selects at random one of the distinct digit-to-letter assignments in the model agent and assimilates it. In our example, the distinct assignments occur for the letters  $A$, $B$, $N$, $O$, $R$ and $T$. Say that letter $B$ is chosen, so that the target agent has to assimilate the assignment $B=3$. To do that the target agent simply interchanges the digits assigned to the letters $B$ and $R$, as in the
elementary move, so that the resulting assignment becomes  $A=0$, $B=3$, $D=9$, $E=4$, $G=8$, $L=1$, $N=7$, $O=6$, $R=2$, $T=5$ whose cost is $c = 1545613$. As expected, the result of  imitation is the increase of the similarity between the target and the model agents, which may not necessarily lead to  a decrease of the cost of the target agent, as in our example.  The case $p=0$ corresponds to the baseline situation where the $M$ agents explore the state space independently.

 It is important to note that in the case the target agent is identical to the model agent, and this situation is not uncommon since the imitation process reduces the diversity of the group, the 
target agent executes the elementary move with probability one. This procedure is different from that used in \cite{Fontanari_14}, in which agents identical to the model agent are not updated   in the imitation action.  Both implementations yield qualitatively similar results, except in the regime where imitation is extremely frequent, i.e., for $p \approx 1$. In particular, for $p=1$ the implementation in which the model agent  is unchanged results in the search being permanently stuck  in a local minimum \cite{Fontanari_14}, whereas in the implementation in which the model agent executes the elementary move actually leads to the optimal performance for very small groups, as we will show next.  The procedure adopted here was used in most studies  of the imitative learning search \cite{Fontanari_15,Fontanari_17}.

%
\begin{figure}
\centering
\includegraphics[width=0.48\textwidth]{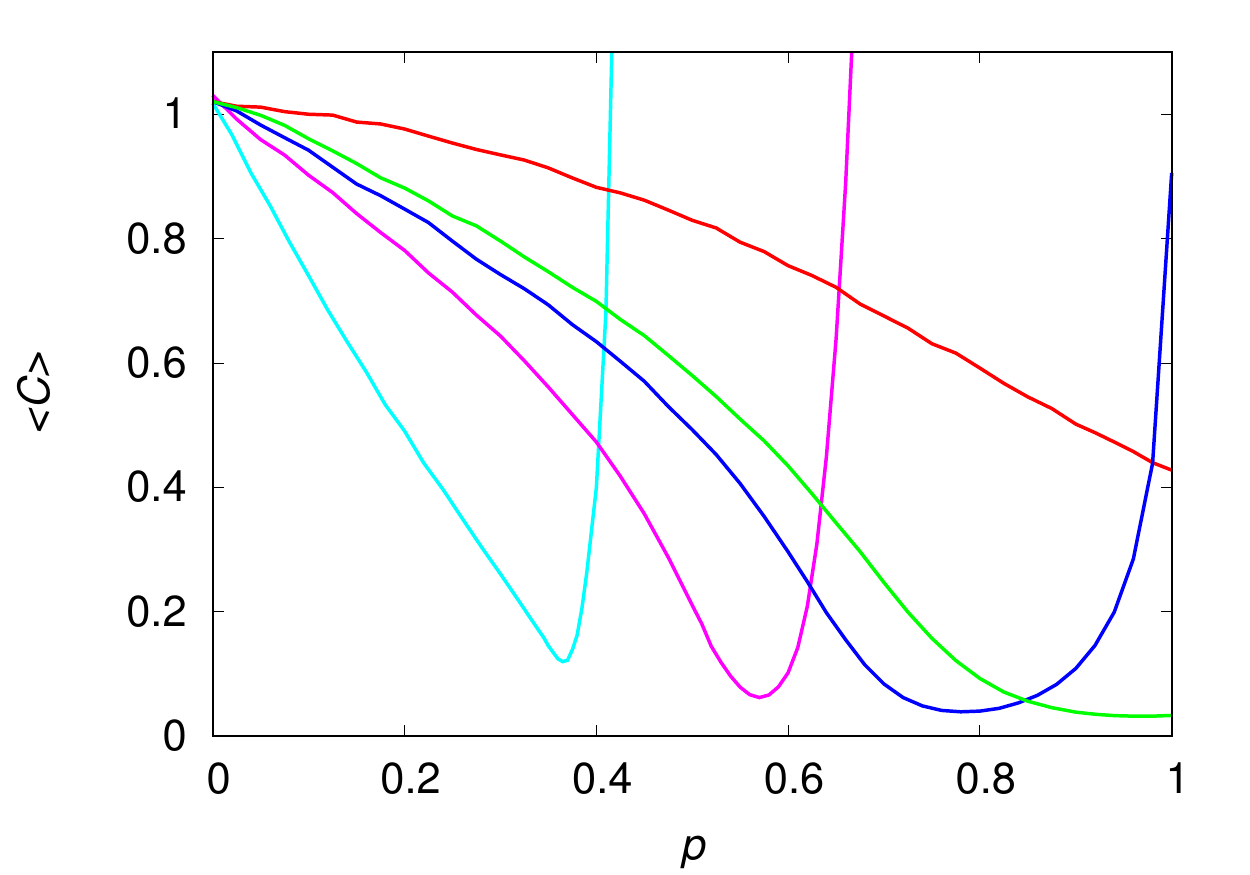}
\caption{Mean computational cost $\langle C \rangle $ 
as function of the  imitation probability  $p$ for groups of size  (top to bottom at $p=0.2$)  $M=2, 5, 8, 25$ and $200$. For the independent search ($p=0$) we find $\langle C \rangle \approx 1.02$ regardless of the group size.
}
\label{fig:I1}
\end{figure}
%

Figure \ref{fig:I1}  shows the mean computational cost $ \langle C \rangle $  as function of the imitation probability  $p$ obtained by averaging over $10^5$ independent runs.  
For groups of size $M < 5$ the performance always improves with increasing $p$   and for those small groups the strategy of always imitating the lowest cost agent (i.e., $p=1$) is optimal.   For $M=5$  there appears a minimum at  $p \approx 0.96$ with computational cost $\langle C \rangle \approx 0.032$ that corresponds to the best performance of the imitative learning search  for the entire space of the model parameters $M$ and $p$. This amounts to more than a thirtyfold improvement on  the group performance as compared with the independent search.
For large groups, increase of the imitation probability $p$  can lead to  catastrophic results due to the trapping of the search around the local minima. This  harmful effect  appears in large groups only  and it is due to the existence of several copies of the model agent carrying a   low cost digit-to-letter assignment (local minimum).   This makes it  very hard to explore other regions of the state space through  the elementary move, since the extra copies attract the updated model agent back to the local minimum.
However, for small group sizes,  the elementary move can easily carry the agents away from the local minima as illustrated in Fig.\  \ref{fig:I1}.  The optimal performance for  $M > 5$, which is determined by the minimum of the curve $ \langle C \rangle $ vs. $p$,  degrades smoothly with increasing $M$. We note that  when the model parameters are close to their optimal values, the distribution of probability of the computational costs is well-described  by  an exponential distribution \cite{Fontanari_14}.

\begin{figure}
\centering
\includegraphics[width=0.48\textwidth]{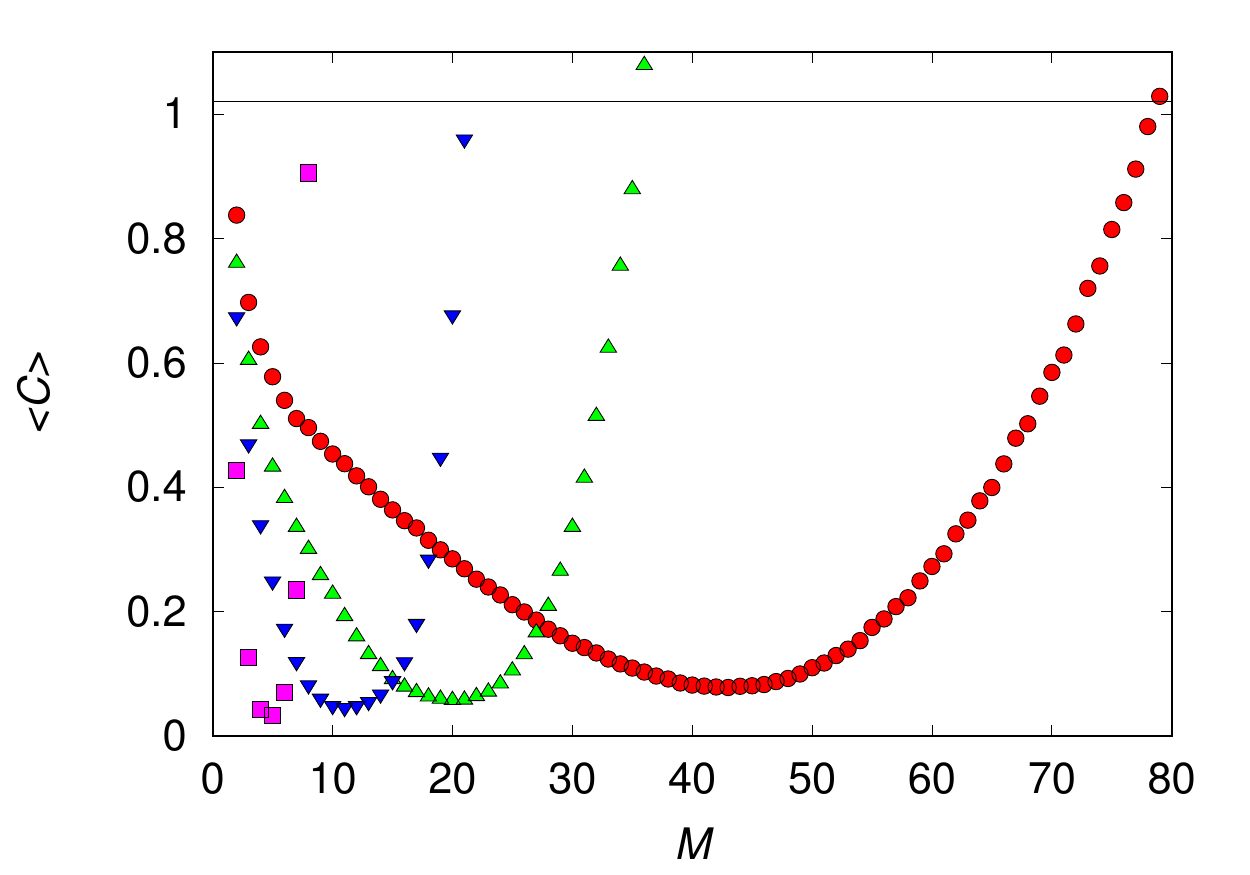}
\caption{Mean computational cost $\langle C \rangle $ 
as function of the  group  size  $M$  for  imitation probability  $p = 0.5$ (circles), $0.6$ (triangles), $0.7$ (inverted triangles) and $1$ (squares).  The horizontal line at $\langle C \rangle = 1.02$ indicates the mean cost for the independent search ($p=0$). 
}
\label{fig:I2}
\end{figure}

Figure \ref{fig:I2}  illustrates more neatly the existence of  a group size that optimizes the performance of the group for a fixed imitation probability.    Hence the conjecture that
the efficacy of imitative learning could be a factor determinant of the group size of social animals \cite{Fontanari_14} (see \cite{Wilson_75,Dunbar_92} for a discussion of the  standard selective pressures on group size in nature).  As pointed out before, the best performance overall is achieved for small groups with  high imitation probability.

%
%

The relevant finding revealed in Figs.\  \ref{fig:I1} and \ref{fig:I2} is that the group performance can be optimized by tuning the two parameters of the  
model, namely, the group size  $M$ and the imitation probability $p$. Hence collective intelligence can be  augmented through a judicious choice of
the behavioral characteristics of the group members, modeled  here by the imitation probability $p$,  and  of the group size $M$.   As already pointed out, the  catastrophic performance  exhibited by  large groups and high imitation probabilities   is akin to the Groupthink phenomenon \cite{Janis_82},  when everyone in a group starts thinking alike,  which can occur when people put unlimited faith in a talented leader (the model agent, in our case).  In addition to using small group sizes, there are  two other ways to avoid the trapping in the local minima. The first way is decreasing or delaying the influence of the model agent by reducing the connectivity of the network \cite{Rodrigues_16}. The second is allowing some diversity in the imitation probabilities of the agents since agents characterized by $p \approx 0$ will rarely be trapped in the local minima \cite{Fontanari_16}.  Although these solutions are effective in avoiding the catastrophic performance for large $M$ and $p$ they have the unwanted side effect of degrading the optimal performance, which is obtained using a fully connected influence network with  homogeneous  imitation probabilities, as in the model described before.

We note that, despite some superficial similarities, the imitative learning search is markedly different from the well-known genetic algorithms \cite{Goldberg_89}. In fact, the elementary move can be seen as  the counterpart of mutations with the caveat that in the genetic algorithm mutation is an error of the reproduction process, whereas in the imitative search the elementary move and the imitation procedure are mutually exclusive processes. The analogy between the imitation and the  crossover processes is  even more far-fetched. The model agent is a mandatory parent in all mates but it contributes  a single gene (i.e.,  a single digit to letter assignment) to the offspring which then replaces the other parent, namely, the target agent. Since the contributed gene is not random - it must be absent in the target agent - the genetic  analogy is clearly inappropriate and so
the imitative learning search stands on its own as a search strategy.

\section{The blackboard organization}\label{sec:black}

The  popular view of  working groups as  teams of specialists   that  exchange ideas on possible approaches to solve a problem and write the promising lines of investigation in a public display  is  the inspiration for  the blackboard organization  \citep{Corkill_91}.  The study of a  minimal model of these brainstorming  groups, which considers $M$  agents and a central blackboard where the agents can read and write hints, 
suggested that the blackboard organization could produce a superlinear speedup of the solution time $t^*$ (see Section \ref{sec:CC}) with respect to the number of group members $M$ \citep{Clearwater_91}. We recall that for the independent search the speedup is linear, i.e., $t^* \propto 1/M$, provided that $M$ is not too large in order to avoid  duplication of work. In our problem,  duplication of work will  occur for unrealistically large groups, $M \gg 10!$, only. If  the superlinear  speedup claim were correct,  then it would offer a nice qualitative evidence of  the benefits of cooperation to problem-solving systems. However, recent evidences indicate that whereas the blackboard organization actually produces a significant improvement on the performance of the search, it does not change the nature of the search which exhibits  the same characteristics of  the independent search and, in particular,  the same scaling of  $t^*$  with $M$
\cite{Fontanari_18}.

An advantage of  blackboard systems is that they do not  need the introduction of  arbitrary  cost functions to weight the quality of
the digit-to-letter assignments. As it will be clear in our  analysis of limited space blackboards,  the number of hints exhibited by an assignment is an effective, albeit indirect, measure of its quality.  A limited space  blackboard  can exhibit at most $B$ hints so that, when the blackboard is full,   the agents must erase  hints to make room for  their own hints  on the board. Next we describe the dynamics of the  blackboard organization. 

At the initial time, $t=1$, all agents' digit-to-letter assignments are  selected with equal probability from the pool of the $10!$ valid assignments. Each agent then checks for all possible hints of its digit-to-letter assignment (see Section \ref{sec:CP}), singles out  the novel hints (i.e., the hints that are not  already displayed on the blackboard) and chooses one  of them at random to  post  on the board.  In this process, the agent also makes a list of the hints that are displayed on the board and that  do not appear in its digit-to-letter assignment. Those are the different hints.
Let us assume that the  limited space board of size $B$ is full,  so the agent must make room to post its selected hint.   To do so the agent  selects one of the different hints on the board at random and replaces it by its hint. In case the board is not full, the agent simply posts the selected hint on the board.  We have tested many variants  of this pick-and-replace procedure   and found that they produce only  negligible quantitative changes on the group performance and so do not affect our conclusions.

Once the initial states of the agents and of the blackboard are set up, the agents can update their digit-to-letter  assignments by  performing two actions: the elementary move described in Section \ref{sec:CP} and the assimilation of one of the hints displayed on the blackboard.
The search procedure develops as follows. It begins with  a randomly chosen agent -- the target agent -- picking a  hint at random from the  blackboard. In the case that there are no hints (i.e, the blackboard is empty), or that the target agent is already using the chosen hint,  the  agent performs the elementary move; otherwise it assimilates the hint.
The assimilation  of a hint by the target agent involves the relocation of at most six digits of its digit-to-letter assignment. 
For example, consider the assimilation of the hint  $\left ( N=1, R=4, B=5 \right )$ by an agent that has the assignment  $A=0$, $B=2$, $D=9$, $E=4$, $G=8$, $L=1$, $N=7$, $O=6$, $R=3$, $T=5$. This can be done sequentially using the same assimilation procedure of the imitative learning search described in Section \ref{sec:imit}. First,   the assignment $N=1$ is assimilated, yielding $A=0$, $B=2$, $D=9$, $E=4$, $G=8$, $L=7$, $N=1$, $O=6$, $R=3$, $T=5$, then $R=4$, yielding $A=0$, $B=2$, $D=9$, $E=3$, $G=8$, $L=7$, $N=1$, $O=6$, $R=4$, $T=5$
and finally $B=5$ resulting in the  digit-to-letter assignment  $A=0$, $B=5$, $D=9$, $E=3$, $G=8$, $L=7$, $N=1$, $O=6$, $R=4$, $T=2$  that exhibits  the desired hint. As usual, after the target agent  is updated, we increment the time $t$ by the quantity $\Delta t = 1/M$.

After any of the events -- elementary move or assimilation of a hint from the blackboard -- the target agent checks for all possible hints from its new assignment and executes the pick-and-replace procedure described before.  In addition, if the solution of the puzzle is found the search halts and the time $t = t^*$ is recorded. 

%
\begin{figure}
\centering
\includegraphics[width=0.48\textwidth]{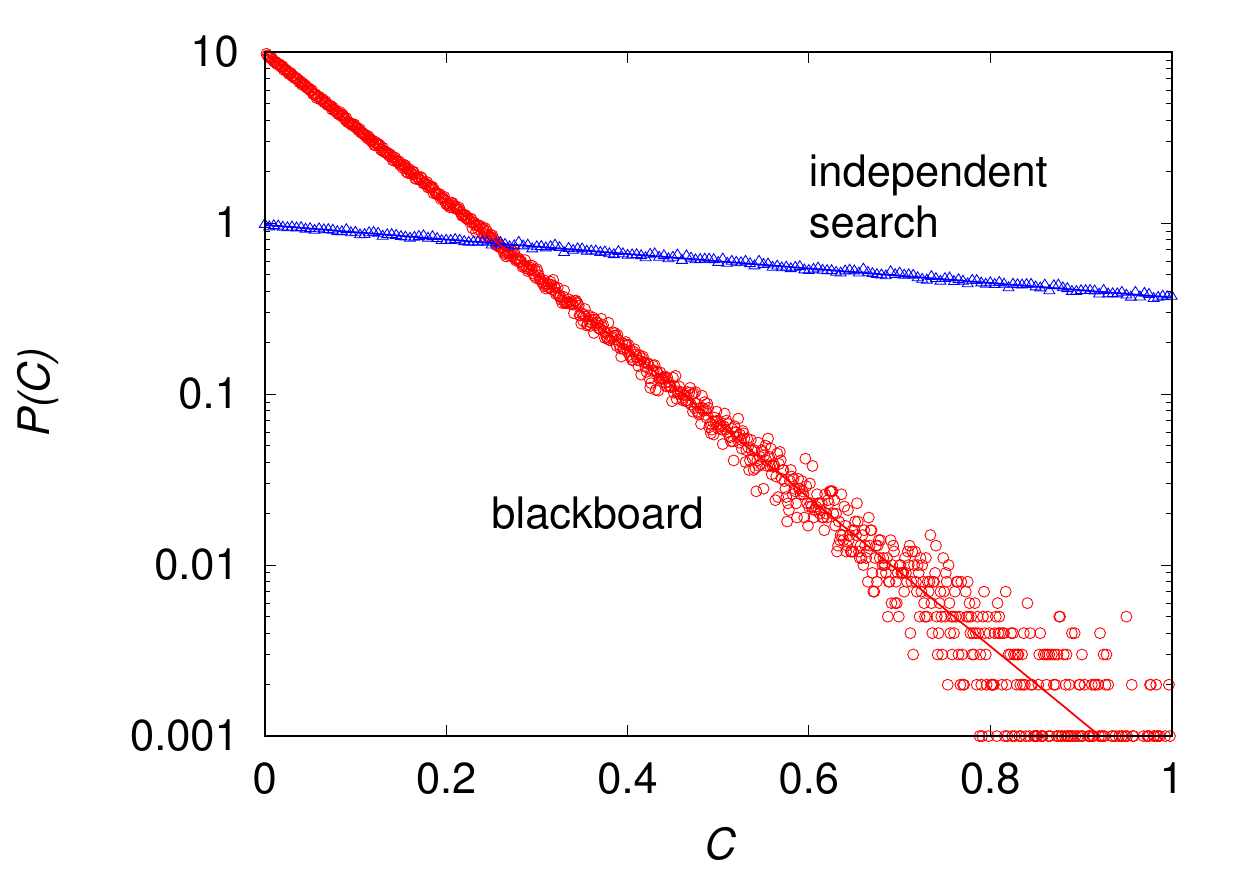}
\caption{Probability distribution of the computational cost (\ref{CC})   for  the independent search and  the  blackboard organization with   $M=100$  agents. There is no space limitation on the blackboard, i.e., $B=351$. These distributions were generated using $10^6$ independent runs.
The curve fitting the data of the independent search is $P \left ( C  \right ) = 0.98 \exp \left ( - 0.98  C \right )$, whereas the data of the
blackboard system is fitted by $P \left ( C  \right ) = 10 \exp \left ( - 10 C \right )$.   }
\label{fig:B1}
\end{figure}
%

Figure \ref{fig:B1} shows the  distribution of probability $ P \left ( C \right ) $ of  the computational cost  for the independent search and for the blackboard organization in the case there is no space limitation on the blackboard so it can display all  351 distinct  hints. 
Our results show that those distributions are exponential for both search strategies, contrary to the suggestion of  \cite{Clearwater_91} that the  exponential distribution, which characterizes the independent search, would be replaced by a lognormal distribution for the blackboard organization.
From a quantitative perspective, however, the  blackboard organization produces a tenfold decrease of the computational cost as compared with 
the independent search.  In particular, $\langle C \rangle \approx 0.10$ for the unlimited space  blackboard and  $\langle C \rangle \approx 1.02$ for the independent search. We note that the  elementary move is  slightly less efficient to explore the state space  than the  replacement of the entire digit-to-letter  assignment (global move) used in  Refs.\ \cite{Clearwater_91,Fontanari_18}. This is  so because it is not too unlikely to reverse a change made by the elementary move. For example, the probability to reverse a change in a subsequent trial is $2/90$ for the the elementary move, whereas it is $1/10!$ for the global move. Interestingly, the replacement of the global by the elementary move has no discernible effect on the performance of 
the blackboard organization.  

%
\begin{figure}
\centering
\includegraphics[width=0.48\textwidth]{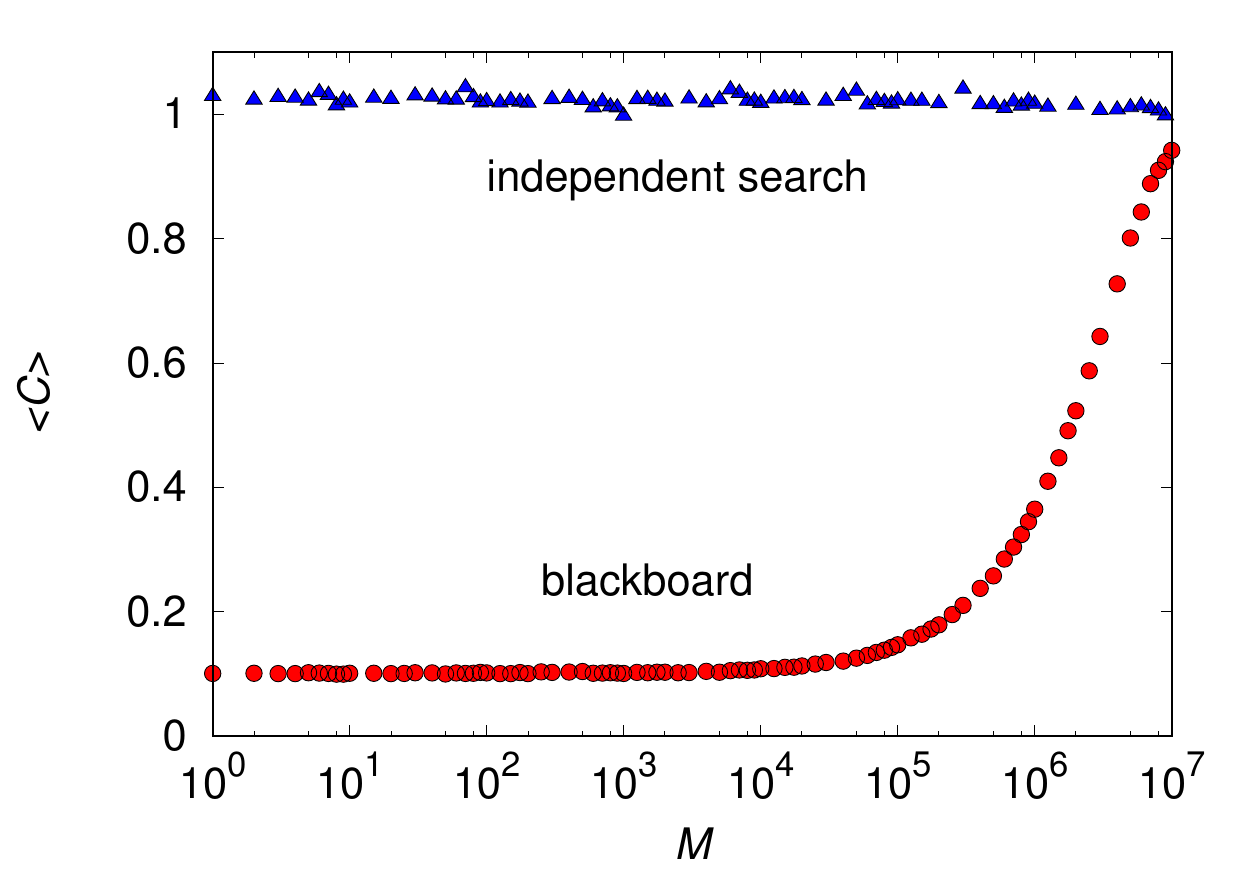}
\caption{Mean computational cost $\langle C \rangle $   as function of the system size $M$ for the independent search (triangles) and the blackboard organization (circles). There is no space limitation on the blackboard, i.e., $B=351$. Each symbol represents the average over $10^5$ independent runs. The error bars are smaller than the symbol sizes.  }
\label{fig:B2}
\end{figure}
%

 Figure \ref{fig:B2}, which shows the mean computational cost as function of the system size, proves that $t^*$ scales with $1/M$  
 (and hence $\langle C \rangle $ is independent of the system size $M$) for both the blackboard  and the independent search strategies. 
 For the blackboard organization, the increase of the computational cost due to duplication of work occurs for much  smaller group sizes than for
 the independent search since the blackboard reduces the effective size of the state space to be explored by the agents.
 The main point is that  the  unlimited  blackboard  system is not really a cooperative problem-solving system, since once the blackboard is filled out, which happens in a very short time \cite{Fontanari_18}, the agents will pick  hints on the board and  explore the state space independently of each other.

%
\begin{figure}
\centering
\includegraphics[width=0.48\textwidth]{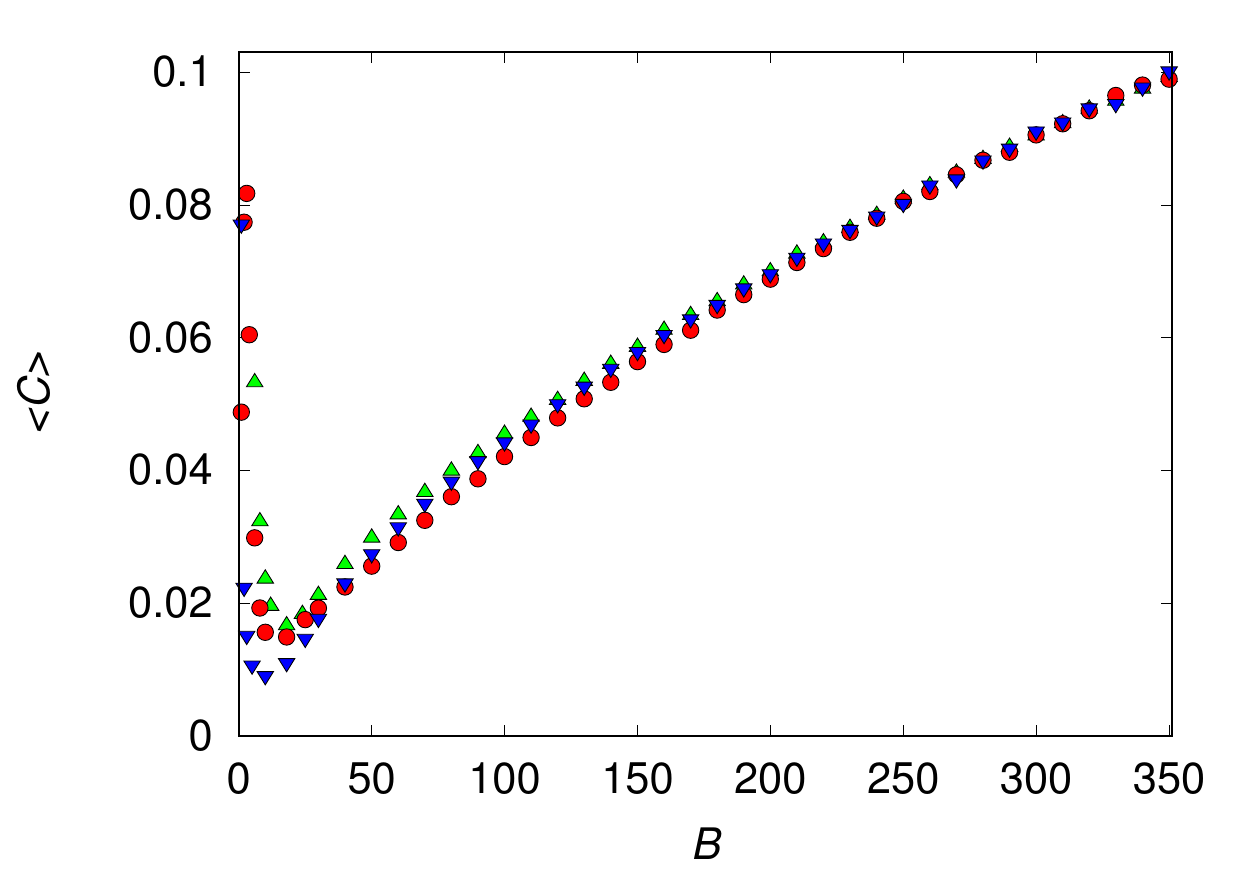}
\caption{Mean computational cost $\langle C \rangle $   as function of the blackboard size $B$ for  systems of size $M=1$ (triangles), $10$ (inverted triangles) and $100$  (circles).  Each symbol represents the average over $10^5$ independent runs. The error bars are smaller than the symbol sizes.  }
\label{fig:B3}
\end{figure}
%

This situation changes dramatically when the size $B$ of the blackboard  is limited so the agents have to compete for space to write their hints on the board. At first sight, one would expect that
limiting the information available to the agents would harm the group performance.  However, Fig.\ \ref{fig:B3} shows that, except for small board sizes ($B < 20$ for the data shown in the figure),
increasing  the number of hints displayed on the board actually  degrades the group performance, regardless of the group size.  In addition, for each group size there is a  value of  $B$ that minimizes the computational cost. In particular, for $M=1$ we find this optimum at $B = 20$, for $M=10$ at $B=7$ and for $M=100$ at $B=15$.  

\begin{figure}
\centering
\includegraphics[width=0.48\textwidth]{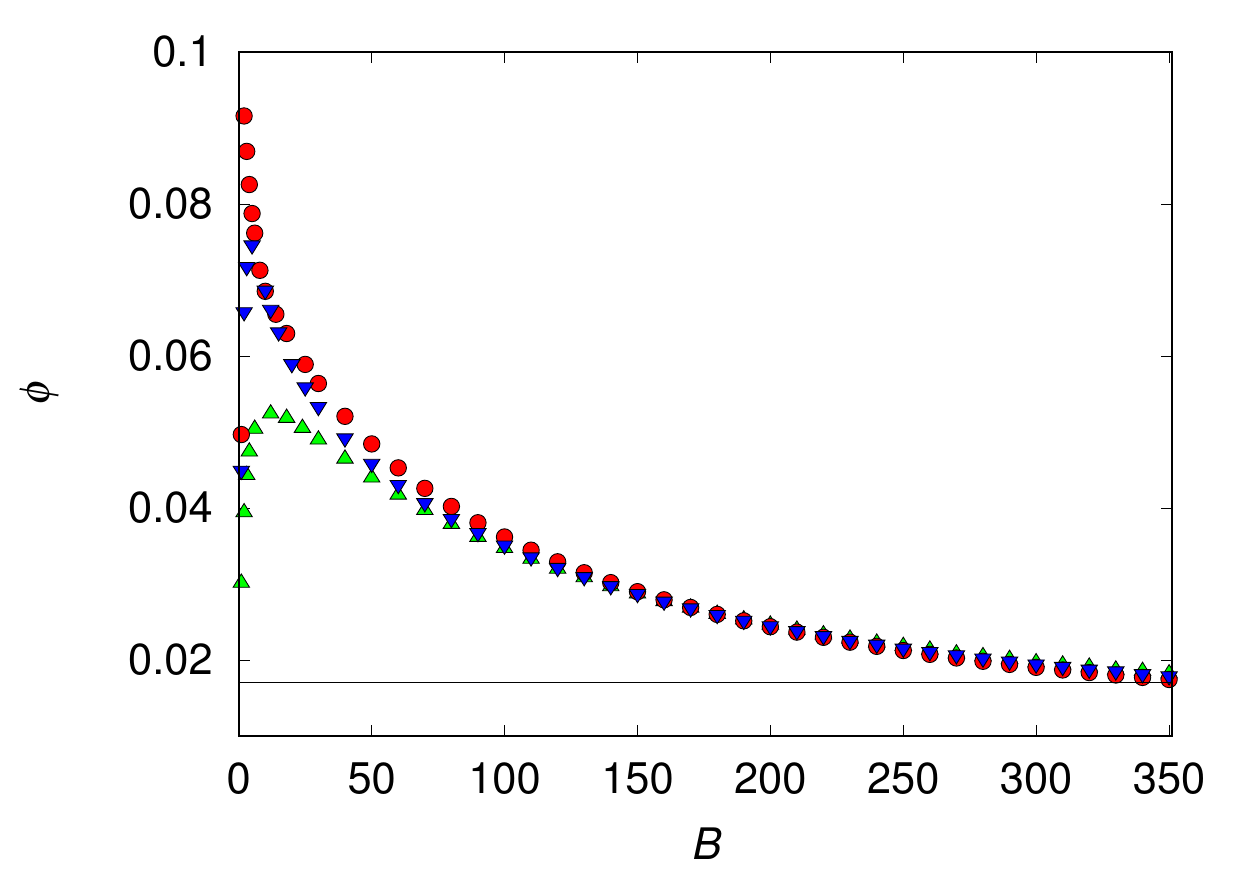}
\caption{Probability $\phi$  that an agent selects a correct hint from the blackboard    as function of the blackboard size $B$ for  systems of size $M=1$ (triangles), $10$ (inverted triangles) and $100$  (circles).  Each symbol represents the average over $10^5$ independent runs. The error bars are smaller than the symbol sizes. The horizontal line at $\phi = 6/351$ is the prediction of the random blackboard null model.  }
\label{fig:B4}
\end{figure}

To understand the counterintuitive finding that limiting  the amount of information displayed on the blackboard improves the group performance,  we present in Fig.\ \ref{fig:B4} the probability $\phi$ that an agent selects a correct hint from the board. This probability is defined as the ratio between the number of correct hints selected and the total number of hint selections,  averaged over all agents during a run. The result is then averaged over $10^5$ independent runs. Not surprisingly, Figs.\ \ref{fig:B3} and \ref{fig:B4} reveal the  strong correlation between  $\langle C \rangle$ and $\phi$ so that the better performance of limited space blackboards is consequence  of  the higher odds  of selecting a correct hint from the board.  But the reason these odds are higher for limited boards is not  obvious at all. For instance, consider a null model in which the hints displayed on the board of size $B$ are selected randomly without replacement from the pool of 351 hints. Since the cryptarithmetic puzzle (\ref{DGR}) has  only six correct hints, the probability  that the board displays exactly $k$ correct hints is given by an hypergeometric distribution.  Now, given that the blackboard displays $k$ correct hints, the   probability that the agents selects one of them is simply $k/B$.
Hence  
the probability that an agent selects a correct hint from a board of size $B$ in this null model is
\begin{equation}\label{null}
\phi = \sum_{k=0}^6 \frac{\binom{6}{k}\binom{351-6}{B-k}}{\binom{351}{B}} \frac{k}{B} = \frac{6}{351} \approx 0.017,
\end{equation}
which does not  depend on the board size. We note that in this null model the  limitation of the number of hints displayed on the board does not affect the group performance at all, contrary to the naive expectation that it would harm that performance. 

The explanation for the dependence on $B$ shown in Fig. \ref{fig:B4} (and consequently in
Fig. \ref{fig:B3}) is that the hints displayed on the board  are not a random sample of the pool of hints, as assumed in the null model. 
The somewhat subtle reason for the bias towards the correct hints is that whenever an agent assimilates a hint from the board it must relocate
up to six digits of its original digit-to-letter assignment. In doing so, it is very likely to eliminate any previous  hints it  carried, except if those
hints, i.e., the hint copied from the board and the hints  that are  already part  of the agent's assignment, are the correct hints. In that sense,  correct hints are insensitive to the radical rearrangement of digits resulting from the assimilation of another correct hint from the board and this explains its
higher frequency in the blackboard.

%
%

As can be hinted from Fig.\ \ref{fig:B3} the dependence of the computational cost on the number of agents $M$  is quite complex and changes qualitatively  for different values of the board size $B$, so we will leave a detailed discussion of this point to a future contribution. Here we mention only  that in a typical situation the computational cost initially decreases with increasing $M$ until it reaches a minimum value, beyond which it begins to increase till it  levels off and becomes size-independent, provided $M$ is not too large,  as in the case of  the unlimited size blackboard (see Fig.\ \ref{fig:B2}).

Finally, the fact that the blackboard organization works so well for just a single agent ($M=1$), which seems to use the blackboard as an external memory to store the hints discovered  during its exploration of the state space,  indicates that  the main role of the blackboard is not the promotion of cooperation between the agents as initially thought \cite{Clearwater_91}: the blackboard serves as a collective memory  storage device for the otherwise  memoryless agents. We note that for the reputation blackboard, where the hints are posted  together with the cost (\ref{cost}) of the agent, the  best performance is achieved in the case of a single agent \cite{Fontanari_18}.  Interestingly, in the context of the wisdom-of-crowds effect \cite{Galton_07} (see also \cite{Surowiecki_04}), the performance of a single individual is usually improved if its estimate is taken as the average of its previous estimates  -- the so-called crowd within  \cite{Vul_08}. These findings in very distinct contexts make  evident the difficulty to disentangle the memory  from the cooperation  effects without the aid of the minimal model approach.

\section{Conclusion}\label{sec:conc}

Rather than  advance new search heuristics to solve combinatorial problems, the goal of our approach to study collective intelligence
 is to assess quantitatively and systematically the potential of cooperation to  solve problems  in very simplified scenarios. Of course, once
 the conditions  that optimize the efficiency of cooperative work are understood, this knowledge  can be used to devise cost-effective search heuristics, which is ultimately the goal of the research on collective intelligence. 
 
 Here we have reviewed and offered original insights on two minimal models of distributed cooperative problem-solving systems, namely, the imitative learning search strategy and the blackboard organization. A good criterion to determine whether a mathematical  model is minimal or not is  the number of model parameters. The imitative learning model has two parameters only: the copy or imitation propensity $p$ of the agents and the group size $M$. The blackboard organization  has two parameters too: the group size $M$ and the blackboard size $B$.  The value of  a minimal  model  (or, for that matter, of any model) should be gauged by the unexpectedness of its predictions. In fact, the  optimization of the group performance for a specific group size and the Groupthink-like phenomenon observed in the study of the imitative learning model, as well as the improved performance of the blackboard organization in the case the amount of available information is limited,  
 bear witness to the importance of those models. 
 
 From a quantitative perspective,  we note that the best performance of the imitative search strategy is achieved for $M=5$ and $p=0.96$ and yields the mean computational  cost $\langle C \rangle \approx 0.03$ (see Fig.\ \ref{fig:I1}), whereas the best performance of the blackboard organization is achieved for $M=B=10$  with mean cost $\langle C \rangle \approx 0.01$ (see Fig.\ \ref{fig:B3}). For the sake of comparison, we recall that the mean computational cost of the independent search is $\langle C \rangle \approx 1.02$, regardless of the group size, so  our two cooperative work  scenarios produce a substantial boost on the performance of the group of agents.

Although the two minimal models of distributed cooperative problem-solving systems presented here  are easy to formulate and simulate in a computer, they exhibit some features that preclude any simple analytical approach as,  for instance, the need to select the best digit-to-letter assignment to serve as model for the agents and the non-local effect of assimilating a hint from the blackboard. Perhaps,  these obstacles may yield to   more powerful and sophisticated mathematical tools, such as the  kinetic theory of active particles \cite{Bellomo_09}, so as to make the study of collective intelligence more appealing to the mathematics community \cite{Burini_16}.

To conclude, we note that the key issues that motivated the proposal of  the original minimal model of cooperative problem-solving systems   remain unanswered \cite{Clearwater_91}. For instance, the question whether cooperative work can alter the  statistical signature of the search on the state space of the combinatorial problem is still open, since the computational costs of the models  studied here are distributed by  exponential probability distributions as in the case of the independent search (see Fig.\ \ref{fig:B1}). Moreover, for large group sizes $M$  the time $t^*$ to find the solution decreases with  $1/M$ for both the blackboard organization  and the independent search (see Fig.\ \ref{fig:B2}), whereas  it actually increases with increasing $M$ for the imitative search due to the Groupthink phenomenon   (see Fig.\ \ref{fig:I2}).   A  qualitative beneficial effect of cooperation should  result in the scaling $t^* \propto 1/M^{\alpha}$ with the exponent $\alpha > 1$, but producing a model with this attribute has proved an elusive task so far.

\acknowledgments
The research  of J.F.F. is  supported in part by grants
2017/23288-0, Fun\-da\-\c{c}\~ao de Amparo \`a Pesquisa do Estado de S\~ao Paulo 
(FAPESP) and  305058/2017-7, Conselho Nacional de Desenvolvimento 
Cient\'{\i}\-fi\-co e Tecnol\'ogico (CNPq). S.M.R. is supported by FAPESP scholarship 2015/17277-0 and
 A.C.A is supported by a CAPES scholarship.

\end{document}